\documentclass{amsart}
\usepackage{amssymb}
\usepackage{verbatim}

\newtheorem{thm}{Theorem}[section]

\newtheorem{lem}[thm]{Lemma}

\newtheorem{cor}[thm]{Corollary}

\newcommand{\cE}{{\mathcal E}}

\newcommand{\Tr}{{{\bf Tr}}}

\newcommand{\bM}{{{\bf M}}}

\newcommand{\A}{{\mathcal A}}
\newcommand{\cB}{{\mathcal B}}

\newcommand{\cM}{{\mathcal M}}
\newcommand{\Cs}{{{$\hbox{\bf C}^*$}}}
\newcommand{\F}{{\mathcal F}}
\newcommand{\Flim}{{{\lim _{\F_0}\, }}}

\newcommand{\cL}{{\mathcal L}}

\newcommand{\cH}{{\mathcal H}}

\newcommand{\Z}{{Z\!\!\! Z}}
\newcommand{\Zd}{{\Z^d}}

\newcommand{\Rn}{{\rm I\!R}}

\newcommand{\cS}{{\mathcal S}}
\newcommand{\I}{1\!{\mathrm l}}

\begin{document}

\title{Quantum $L_p$ and Orlicz Spaces}

\author{Louis E. Labuschagne}
\address{Department of Maths, Applied Maths and Astronomy,
P.O.Box 392, University of South Africa, 0003 Pretoria, South
Africa} 
\email{labusle@unisa.ac.za}
\author{W{\l}adys{\l}aw A. Majewski}
\address{Institute of Theoretical Physics and Astrophysics, Gda{\'n}sk
University, Wita Stwo\-sza~57, 80-952 Gda{\'n}sk, Poland} 
\email{fizwam@univ.gda.pl}

\begin{abstract}
Let $\A$ ($\cM$) be a $C^*$-algebra (a von Neumann
algebra respectively).
By a quantum
dynamical system we shall understand the pair $({\A} ,T)$ 
($({\cM}, T)$) where $T : {\A} \to {\A}$ ($T : {\cM} \to {\cM}$)
is a linear, positive (normal respectively), and identity
 preserving map.
In our lecture,
we discuss how the techniques of
 quantum Orlicz spaces may be used to study 
quantum dynamical systems. 
To this end, we firstly give a brief exposition of the theory of quantum dynamical systems in quantum $L_p$ spaces. Secondly,
we describe the Banach space approach to quantization of classical Orlicz spaces.
We will discuss the necessity of the generalization of $L_p$-space
techniques. Some emphasis will be put on the construction of non-commutative Orlicz spaces.
The question of lifting dynamical systems defined on von Neumann algebra to a dynamical system defined in terms of quantum Orlicz space will be discussed.
\end{abstract}



\maketitle

\section{Introduction}

To indicate reasons why (quantum) $L_p$-spaces are emerging in the theory of (quantum) dynamical systems we begin with a particular case of dynamical systems - with stochastic evolution of particle systems. We recall that
in the classical theory of particle systems one of the objectives is to produce, describe, 
and analyze dynamical systems
with evolution originating from stochastic processes in such a way that their 
equilibrium states are given Gibbs states  (see
\cite{L}). 
A well known illustration is the so called Glauber dynamics \cite{G}, which may be found in 
a number of papers. To carry out the analysis of 
such dynamical systems, it is convenient 
to use the theory of Markov processes in
the context of $L_p$-spaces. 
In particular, for the Markov-Feller processes, using the unique correspondence
between the process and the corresponding dynamical semigroup, one can give a recipe 
for the construction of Markov generators for this class of processes (for details see \cite{L}).
That correspondence uses the concept of conditional expectation which can be nicely
characterized within the $L_p$-space framework (cf. the Moy paper \cite{Moya}). 

More generally,  these Banach spaces, i.e.
$L_p$ and their generalizations - Orlicz spaces,
are extremely useful in the general description of classical dynamical systems. To support this claim
some comments are warranted here. Firstly
let $\{\Omega, \Sigma, \mu\}$
be a probability space. We denote by
 ${\cS}_{\mu}$  the set of the densities of all the probability measures equivalent to $\mu$, i.e.,

$${\cS}_{\mu} = \{ f  \in L^1(\mu): f>0 \quad \mu-a.s., E(f) =1 \}$$

${\cS}_{\mu}$ can be considered as a set of (classical) states and its natural ``geometry''  comes from embedding ${\cS}_{\mu}$ into $L^1(\mu)$. However, it is worth pointing out that the 
Liouville space technique demands $L^2(\mu)$-space, while 
employing the interpolation techniques needs other $L_p$-spaces with $p \geq 1$.

To take one further step, let us consider 
moment generating functions; so fix $f \in {\cS}_{\mu}$ and take a real random variable $u$ on $(\Omega, \Sigma, f d\mu)$.
Define

$$\hat{u}_f(t) = \int exp(tu) f d\mu, \qquad t \in \Rn $$

and denote by $L_f$  the set of all random variables such that
\begin{enumerate}
\item{} $\hat{u}_f$ is well defined in a neighborhood of the origin $0$,
\item{} the expectation of $u$ is zero.
\end{enumerate}
One can observe that in this way a nice selection of (classical) observables was made, namely \cite{PS}
 all the moments of every $u \in L_f$ exist and they are the values at $0$ of the derivatives of $\hat{u}_f$.

 But, it is important to note that $L_f$ is actually the {\bf Orlicz space} based on an exponentially growing function (see  \cite{PS}).
Consequently, one may say that even in classical statistical Physics one could not restrict oneself to merely $L^1(\mu)$, $L^2(\mu)$, $L^{\infty}(\mu)$
and interpolating $L^p(\mu)$ spaces. In other words,  generalizations of $L_p$-spaces - Orlicz spaces - do appear.

However, contemporary science has been founded on {\it quantum mechanics}. Therefore, it is quite
natural to look for the quantum counterpart of the above approach. Again let us begin with a particle systems with a stochastic evolution.
Recently, the quantization of such particle systems was carried out, see \cite{R7,R8,R9,R10}. 
The main ingredient of such a quantization, is the concept of a generalized conditional expectation and 
Dirichlet forms defined in terms of {\it non-commutative (quantum) $L_p$-spaces}.
The advantage of using quantum $L_p$-spaces, lies in the fact that when
performing the quantization procedure, we can 
follow the traditional ``route'' of analysis of dynamical systems, and also in the fact 
that it is then possible to have one scheme for the quantum counterparts of stochastic dynamics of
jump and diffusive-type. In particular,  the quantum counterpart of the
classical recipe for the construction of quantum Markov generators was obtained.
The above scheme is not surprising if we realize that even in the textbook formulation of Quantum Mechanics, states are trace class operators. So, they form a subset of quantum $L_1(\cB(\cH)), Tr)$-space while observables can be identified with self-adjoint elements of $L_{\infty}(\cB(\cH), Tr)$-space.

Turning to quantum Orlicz spaces our first remark is that they are a natural generalization of $L_p$ spaces.
To provide a simple argument in favor of such a generalization we will follow Streater \cite{S1,S2} . Let $\varrho_0$
be a quantum state (a density matrix) and $S(\varrho_0)$ its von Neumann entropy. Assume $S(\varrho_0)$
to be finite. It is an easy observation that in any neighborhood of $\varrho_0$ (given by the trace norm, so in the 
sense of quantum $L_1$-space) there are plenty of states with infinite entropy. This should be considered alongside the thermodynamical rule which tells us that the entropy should be a state function which is increasing in time. Thus we run into serious problems with the explanation of the phenomenon of return to equilibrium. More sophisticated arguments in this direction can be extracted from  hypercontractivity of quantum maps and log Sobolev techniques (see \cite{OZ} and B. Zegarlinski lecture in \cite{AF}.) 

The paper is organized as follows: in Section 2 we review some of the standard facts on quantum spin systems. Then quantum $L_p$-spaces are described (Section 3). In Section 4, we indicate how $L_p$-space techniques can be used for the construction of quantum stochastic dynamics. Section 5 is devoted to the study of quantum Orlicz spaces.

We want to close this section with a note that the quantum $L_p$ space technique ``ideology'', presented here, is reproduced from the paper \cite{Maj}
which, to some extent, due to technical problems, is unreadable. 

\section{Quantum spin systems}

In this Section we recall the basic elements of the description of
quantum spin systems on a lattice.
The best general references are \cite{BR, Ru}. 
Here, and subsequently, $\Zd$ stands for the d-dimensional integer 
lattice. Let $\F$ denote the family
of all its finite subsets and let $\F_0$ be an increasing Fisher (or van Hove)
sequence of finite volumes invading all of the lattice $\Zd$. Given a sequence
of objects $\{ F_\Lambda \}_{\Lambda \in\F_0}$, it will be convenient
to denote its limit (in an appropriate topology) as $\Lambda \to\Z^d$  
through the sequence $\F_0$ by $\Flim F_\Lambda $.

The basic role in the description of the quantum lattice systems, is played
by a \Cs - algebra  $\A $, with norm $||\cdot||$, defined as
the inductive limit over finite dimensional complex matrix algebras $\bM$. 
By analogy with the classical commutative spin systems, it is natural to view 
$\A$ as a noncommutative analogue of the space of bounded continuous functions. 
For a finite set $X\in\F$, let $\A_X$ denote a subalgebra
of operators localised in the set $X$. We recall that such a subalgebra 
is isomorphic to $\bM^X$. For an arbitrary subset $\Lambda\subset\Zd$, one
defines $\A_\Lambda $ to be the smallest (closed) subalgebra of $\A$ 
containing $\bigcup\{ \A_X : X\in\F,\, X\subset \Lambda\} $.   
An operator $f\in\A$ will be called local if there
is some $Y\in\F$ such that $f\in\A_Y$. 
The subset of $\A$ consisting of all local operators will be denoted by
$\A_0$. 
(A detailed account of matricial and operator algebras can be found in \cite{KR}.)

Together with the algebra $\A$, we are given a family  
$\Tr_X$, $X\in\F$, of {\em normalised partial traces} on $\A$.
We mention that the partial traces $\Tr_X$ have all the natural properties
of classical {\em conditional expectations}, i.e. they are (completely) 
positive, unit preserving projections on the algebra $\A$.
There is a unique state $\Tr $ on $\A$, called
{the normalised trace}, such that
\begin{equation}
\Tr\left(  \Tr_X f\right) =
\Tr\left(f\right)
\end{equation}
for every $X\in\F$, i.e. the normalised trace can be regarded as 
a (free) Gibbs state in a similar sense as in classical statistical 
mechanics.

To describe systems with interactions, we need to introduce the notion of
an interaction potential.  
A family $\Phi \equiv \{ \Phi_X\in\A_X\}_{X\in\F}$ of selfadjoint operators 
such that
\begin{equation}
\Vert \Phi \Vert_1 \equiv \sup_{i\in\Zd} \sum_{X\in\F\atop{X\ni i}}
 \Vert \Phi_X\Vert < \infty
\end{equation}
will be called a (Gibbsian) potential. 
A potential $\Phi \equiv \{ \Phi_X \}_{X\in\F}$ is
of {\it finite range} $R\ge 0$, iff 
$ \Phi_X  =0$ for all $X\in \F,\quad diam(X) >R$.
The corresponding Hamiltonian $H_\Lambda$ is defined by
\begin{equation}
H_\Lambda \equiv H_\Lambda (\Phi )\equiv \sum_{X\subset\Lambda} \Phi_X
\end{equation}
In particular, it is an easy observation that {\it anisotropic and isotropic Heisenberg models (so also
Ising model) with nearest-neighbor interactions fall into the considered class 
of systems!}

Using the Hamiltonian $H_{\Lambda}$, we introduce a density matrix $\rho_\Lambda$ 
$$
\rho_\Lambda \equiv \frac {e^{-\beta H_\Lambda}}  {\Tr e^{-\beta H_\Lambda}}
$$
with $\beta \in (0,\infty)$, and 
define a finite volume Gibbs state $\omega _\Lambda $ as follows:
$$
\omega_\Lambda (f)\equiv \Tr \left( \rho_\Lambda f\right)
$$
It is known, see e.g. \cite{BR}, that for 
$\beta\in(0,\infty)$ 
the thermodynamic limit state on $\A$ 
\begin{equation}
\omega \equiv \Flim  \omega_\Lambda
\end{equation}
exists and is  faithful for some exhaustion $\F_0$ of the lattice.
In general, a system can possess several such states, so phase transitions are allowed.
For a quantum spin system, 
we can also introduce a natural Hamiltonian dynamics
defined in a finite volume as the following automorphism group 
associated with the potential $\Phi $:
\begin{equation}
\alpha_t^\Lambda(f) \equiv e^{+it H_\Lambda} f e^{-it H_\Lambda}
\end{equation}
If the potential $\Phi \equiv \{ \Phi _X\}_{X\in\F}$ also satisfies
\begin{equation}
\label{war1}
||\Phi||_{exp }
\equiv 
\sup_{i\in\Zd}\sum_{X\in\F\atop{X\ni i}} e^{\lambda|X|}||\Phi_X||
<\infty
\end{equation}
for some $\lambda >0$, then the limit 
\begin{equation}
\alpha_t (f) \equiv \Flim  \alpha_t^\Lambda (f),
\end{equation}
exists \cite{BR} for every $f\in\A_0$.
Consequently, the specification of local interactions, leads
to a well defined global dynamics, provided that (\ref{war1}) is valid.
In other words, the thermodynamic limit 
$$
(\A_{\Lambda},\alpha_t^{\Lambda}, \omega_{\Lambda} )
\to (\A, \alpha_t, \omega)
$$ 
exists and gives the quantum dynamical system.

\section{Non-commutative $L_p$-spaces.}

Let $<X,\mu>$ be a measure space, and $p\ge1$. We denote by
$L_p(X,d\mu)$ the set (of equivalence classes) of measurable 
functions satisfying
$$\Vert f \Vert_p \equiv \Bigl( \int_X \vert f(x) \vert^p
d\mu(x)\Bigl)^{\frac{1}{p}} < \infty.$$
For the pair $(\mathcal M, \tau)$ consisting of semifinite von 
Neumann algebra $\mathcal M$ and a trace $\tau$, the analogue of 
the concept of $L_p$-spaces  
($p\in[1,{\infty}]$) in the commutative theory, 
can be introduced as follows : define
$${\mathcal I}_p=\{ x \in {\mathcal M} \quad \vert \quad 
{\tau}(\vert x \vert^p) < {\infty} \}.$$
${\mathcal I}_p$ is a two sided ideal of $\mathcal M$. Further, 
$\Vert x \Vert_p = \tau(\vert x \vert^p)^{\frac{1}{p}}$ 
defines a norm on ${\mathcal I}_p$. The completion of ${\mathcal I}_p$ 
with respect to the norm $\Vert \cdot \Vert_p$ gives Banach 
$L_p(\mathcal M,\tau)$ spaces which can be considered 
as a generalization of the corresponding spaces defined in the commutative case. 
It is an easy observation that on setting ${\mathcal M} = {\mathcal B}(\cH)$
and $\tau = Tr$ ($Tr$ stands for the usual trace on $\mathcal M$), 
one obtains the well known Schatten classes \cite{Schat} . 
That is, $L_p({\mathcal B}(\cH),Tr)$ is just the set of compact operators whose singular
values are in $l_p$ and the norms of $L_p$ and $l_p$ are equal.
Moreover, the family $\{ L_p({\mathcal B}(\cH),Tr) \}_{p \ge 1}$ provides a nice example
of an abstract interpolation scheme (see \cite{RS}).

Using this and the Haagerup theory
(\cite{Haa}; see also \cite{ArM,Hi,Ko,Ne,Te1,Te2}), we can introduce {\it quantum $L_p$ spaces for quantum lattice systems},
i.e. for the systems described in the previous Section.

To this end, we firstly note that the quasi-local structure
described for quantum lattice systems, 
can be summarized in the following way:
\begin{enumerate}
\item
$\A_0 = \cup_{\Lambda \in \F} \A_{\Lambda}$
is dense in $\A$.
\item There exists a family of density operators 
$\{ \varrho_{\Lambda} \in \A_{\Lambda}: \varrho_{\Lambda} 
>0, \Tr\varrho_{\Lambda}=1 \}_{\Lambda \in \F}$
with the compatibility condition 
$\Tr_{\Lambda_2 \backslash \Lambda_1}  \{ \varrho_{\Lambda_2} \} = \varrho_{\Lambda_1}$,
 provided that $\Lambda_1 \subset \Lambda_2$.
\end{enumerate}

We introduce:
\begin{itemize}
\item $||f||_{L_{p,s}(\omega)} = 
\lim_{\Lambda}||f||_{L_{p,s}(\omega_{\Lambda})}$ 
for $p \in [1, \infty),  s \in [0,1]$, where $f \in \A$,
\item
$||f||_{L_{p,s}(\omega_{\Lambda})} =
 (\Tr|\varrho^{1 -s/p}_{\Lambda}f \varrho^{s/p}_{\Lambda}|^p)^{1/p}.$
\end{itemize}
where 
$\omega(f) = \Flim \omega_{\Lambda}(f)
 \equiv \Flim \Tr\{ \varrho_{\Lambda} f \}$.

One can show that $||f||_{L_{p,s}(\omega_{\Lambda})}$ 
is a well defined two-parameter family of norms on $\A$.
The same should be done for $||f||_{L_{p,s}(\omega)}$ (see Theorem below).

Namely, in \cite{R7,R9} it was proved:

\bigskip

\begin{thm}

For any $p\in[2,\infty)$, $s\in[0,1]$, any local operator
$f\in\A_{\Lambda_0},\, \Lambda_0\in\F$ and all sets 
$\Lambda_1 , \Lambda_2\in\F$ such that 
$\Lambda_0\subset\Lambda_1\subset \Lambda_2$, we have
$$
|| f ||_{{ L}_p(\omega_{(\Lambda_2)},s)} 
\leq || f ||_{{L}_p(\omega_{(\Lambda_1)},s)}. 
$$
Thus for any $f\in\A_0$ the limit
$$
|| f ||_{{L}_p(\omega,s)} \equiv 
\Flim || f ||_{{L}_p(\omega_{(\Lambda)},s)}  
$$
exists and is independent of the countable exhaustion $\F_0$ 
of the lattice.
\end{thm}
\bigskip

For $p \in (1,2)$ one can use duality to define the correspondings norms \cite{R9} :
$$||f||_{{L}_p(\omega,s)} \equiv sup_{||g||_{{L}_q(\omega,s)}\leq1} <g,f>_{\omega,s}$$
where $1/p +1/q = 1$, $q \in [2, \infty)$ and $<\cdot, \cdot>_{\omega,s}$ is the scalar product associated to the norm $||\cdot||_{{L}_2(\omega,s)}$. Finally, the existence of the norm  \newline $||\cdot||_{{L}_1(\omega,s)}$ was established in \cite{R7}.
Hence quantum $L_p$-spaces are associated with concrete physical systems: 
\begin{cor}
To every Gibbs state $\omega$ on a $C^*$-algebra $\A$ 
defined by a quantum lattice system we can 
associate
an interpolating, two parameter, family of Banach spaces 
$$\{{L}_p(\omega,s)\}_{p\in[1,\infty), s\in[0,1]}. $$
\end{cor}

\section{Quantum $L_p$ dynamics}

Let $\cM$ be a von Neumann algebra generated by $\pi_{\omega}(\A)$, where $\pi_{\omega}(\cdot)$ is the GNS representation
associated with the quantum lattice system $(\A, \omega)$, described in Section 2. By $\varphi_1$ we denote the (weak) extension of $\omega$ on $\cM$.
Let $\cE_0$ be a  conditional expectation, i.e.
$\cE_0(f^*f) \ge0$, $\cE_0({\bf 1}) = \bf 1$, $\cE_0^2 = \cE_0$.
We define
\begin{equation}
\varphi_2(\cdot) \; \equiv \; \varphi_1 \circ \cE_0(\cdot).
\end{equation}
Suppose that $\varphi_2$ is another faithful state on $\cM$.
Then the Takesaki theorem implies that $\cE_0$
commutes with $\sigma^2_t$ (the modular automorphism group for $ (\cM,
\varphi_2)$)
and hence is symmetric
in $\bigl( \cH_{2,{\frac{1}{2}}}, <f, g>_{2,{\frac{1}{2}}}  \; \equiv \; \varphi_2((\sigma^2_{\frac{i}{4}}(f)^* 
(\sigma^2_{ \frac{i}{4}}(g))) \bigr) $.

Let $V_t \equiv (D\varphi_1:\,D\varphi_2)_t$ be 
the Radon-Nikodym cocycle. We remind that, in particular,
$\sigma^1_t(f) = V_t^* \sigma_t^2(f) V_t$.
The main difficulty in carrying out the construction
of the Markov generator, is the existence of an analytic extension
of $\Rn \ni t \mapsto V_t \in \cM$.
The following condition guarantees the desired extension (for details see \cite{Conn}): 

Suppose there exists a positive constant $c \in (0, \infty)$
such that for any $0 \le f \in \cM$ the following
inequalities hold:
\begin{equation}
\label{nierow}
{\frac{1}{c}} \varphi_1(f) \le \varphi_2(f) \le c
\varphi_1(f).
\end{equation}
Then, $V_t$ extends analytically to $ -  \frac{1}{2} \le
Im z \le \frac{1}{2}$
and $\xi \equiv V_{t \vert t= - {\frac{i}{2}}}$
is a bounded operator in $\cM$.
Let us note that the above inequalities also
guarantee that $\varphi_2$ is a faithful state
provided that $\varphi_1$ has this property.\par

Now, let us apply the above strategy to a finite system. Fix $X \subset \Lambda \in \F$.
Obviously, (\ref{nierow}) is satisfied for $\varphi_1(\cdot)
( \equiv \varphi_1^{\Lambda}(\cdot)) = \Tr_{\Lambda}\varrho_{\Lambda}(\cdot)$ 
and $\varphi_2(\cdot) ( \equiv \varphi_2^{\Lambda,X}(\cdot)) =
\varphi_1 \circ \Tr_X (\cdot)$.
Define
$$\cE_{X,\Lambda}(a) = \Tr_X( \gamma^*_{X, \Lambda} f \gamma_{X, \Lambda})$$
where $\gamma_{X, \Lambda} = \varrho^{\frac{1}{2}}_{\Lambda}
(\Tr_X \varrho_{\Lambda})^{-\frac{1}{2}}$, and $f \in \A_{\Lambda}$.

One can verify \cite{R10} that $\gamma_{X, \Lambda}$ is the analytic extension of 
the Radon-Nikodym cocycle, and that $\cE_{X, \Lambda}$ is a generalized conditional expectation
(in the Accardi-Cecchini sense). Moreover \cite{R7} ,
$$P^{X, \Lambda}_t \equiv exp\{t( \cE_{X, \Lambda} - {\it id}) \}$$
is the well defined Markov semigroup  corresponding to the block-spin flip operation. For its
construction {\em only local specifications $(\varrho_{\Lambda}, \Tr_X\varrho_{\Lambda})$
are necessary}.

Now we examine (like in the classical case) the question of existence
of global dynamics. 
Denoting $\varphi_2 \equiv \varphi_1 \circ \Tr_X$ and 
using the same strategy, we have \cite{R9}
\begin{thm}
\label{Th1}
{ Suppose the system is in sufficiently high
temperature, $\vert \beta \vert < \beta_0$ with interaction
$\Phi$ fulfilling the condition (\ref{war1}),
or that the system is one dimensional at an arbitrary temperature 
$\beta \in (0, \infty)$ with finite
range interactions. Then, for some positive $c \in (0,\infty)$
$$ {\frac{1}{c}} \varphi_1 (f^*f) \le \varphi_2(f^*f)
\le c \varphi_1(f^*f).$$
Hence, the corresponding
Radon-Nikodym cocyles have analytic extension and therefore
$\gamma_X  \equiv (D\varphi_1:D\varphi_2)_{\vert t= -{\frac{i\beta}{2}}}
\in \cM$. Hence
$$\cE_X(f) \equiv \Tr_X(\gamma_X^* f \gamma_X)$$
defines a generalized conditional expectation
which is symmetric in $\cH_{\varphi_1}$. (Here $\cH_{\varphi_1}$ 
is just the Hilbert space ${L}_2(\varphi_1,1/2)$ constructed on $\cM$)}
\end{thm}

On the other hand, one has (for details see \cite{R9}):

\begin{thm}
\label{Th2}
Let $\cE_0$ be a (true) conditional expectation (so not necessary
of the form $\Tr_X$).
Assume that  $\xi \equiv V_{t \vert t= - {\frac{i}{2}}}$
is a bounded operator in $\cM$ and define
$$\cE(f) \; \equiv \; \cE_0(\xi^* f \xi).$$
Then, the generalized conditional expectation $\cE(\cdot)$
is well defined and it has the following properties:
\begin{enumerate}
\item $\cE({\bf 1}) = {\bf 1},$
\item $\cE(f^*f) \ge 0,$
\item $<\cE(f),g>_1 = <f, \cE(g)>_1.$
\end{enumerate}
where $<f,g>_1 \equiv \varphi_1\bigl((\sigma^1_{\frac{i}{4}}(f))^* 
(\sigma^1_{\frac{i}{4}}(g)) \bigr) $.
\end{thm}
\bigskip
Here, again, the generalized conditional expectations are understood
in the Accardi-Cecchini sense (cf. \cite{A,AC,Ta}).
Thus we arrive at:
\begin{cor} {\em Theorems \ref{Th1} and \ref{Th2}
 ensure that the operator given by:
$$\cL \;\equiv \; \cE - {\it id}.$$
 is  a well defined Markov generator.}
 \end{cor}
Consequently, the (Markov) global quantum stochastic semigroup
$P_t \equiv e^{t \cL}$ can be constructed (for high temperature region). It is worth pointing out that $P_t|_{\cM}$ are completely positive (CP) maps on the von Neumann algebra $\cM$ and bounded with respect to $L_2(\varphi_1,\frac{1}{2})$ norm (see \cite{R7, R9}). So, they give rise to well defined maps on quantum $L_2$-space.  In a similar way, one can perform quantization of other stochastic dynamics \cite{R10,AF} .

\bigskip

However, it is important to note here that we were forced to restrict ourselves to high temperature regions (for 
lattice systems of dimension larger than 1). As we were not able to overcome this difficulty \cite{R11}, one may postulate that besides to the suggestions mentioned in the Introduction, some generalization of quantum $L_p$ spaces could be useful. But to take these hints seriously, one should as a first step study the problem of lifting quantum maps (considered dynamical maps are CP maps on a von Neumann algebra) to well defined maps on quantum Orlicz spaces. This will be done in the next Section. 

\section{Orlicz spaces}

Let us begin with some preliminaries.
By the term an \emph{Orlicz function} we understand a convex function 
$\phi : [0, \infty) \to [0, \infty]$ satisfying $\phi(0) = 0$ and $\lim_{u \to \infty} 
\phi(u) = \infty$, which is neither identically zero nor infinite valued on all of 
$(0, \infty)$, and which is left continuous at $b_\phi = \sup\{u > 0 : \phi(u) < 
\infty\}$.  In particular, any Orlicz function 
must also be increasing. 

Let $L^0$ be the space of measurable 
functions on some $\sigma$-finite measure space $(X, \Sigma, m)$. The Orlicz space 
$L^0_{\phi}{}$ associated with  $\phi$ is defined to be the set $$L^{\phi} = \{f \in 
L^0 : \phi(\lambda |f|) \in L^1 \quad \mbox{for some} \quad \lambda = \lambda(f) > 0\}.$$
This space turns out to be a linear subspace of $L^0$ which becomes a Banach space when 
equipped with the so-called Luxemburg-Nakano norm 
$$\|f\|_\phi = \inf\{\lambda > 0 : \|\phi(|f|/\lambda)\|_1 \leq 1\}.$$
 
Let $\phi$ be a given Orlicz function. In the context of semifinite von Neumann algebras 
$\cM$ equipped with an fns trace $\tau$, the space of all $\tau$-measurable operators 
$\widetilde{\cM}$ (equipped with the topology of convergence in measure) plays the role of 
$L^0$ (for details see \cite{Ne}). In the specific case where $\varphi$ is a so-called Young's function, Kunze \cite{Kun} 
used this identification to define the associated noncommutative Orlicz space to be 
$$L^{ncO}_{\phi}{} = \cup_{n=1}^\infty n\{f \in \widetilde{\cM} : \tau(\phi(|f|) \leq 1\}$$ 
and showed that this too is a linear space which becomes a Banach space when equipped with the 
Luxemburg-Nakano norm $$\|f\|_\phi = \inf\{\lambda > 0 : \tau(\phi(|f|/\lambda)) \leq 
1\}.$$Using the linearity it is not hard to see that $$L^{ncO}_{\phi}{} = \{f \in 
\widetilde{\cM} : \tau(\phi(\lambda|f|)) < \infty  \quad \mbox{for some} \quad \lambda = 
\lambda(f) > 0\}.$$ 
Thus there is a clear analogy with the commutative case.

It is worth pointing out that
there is another approach to Quantum Orlicz spaces. Namely, one can replace $(\cM, \tau)$ by $(\cM, \varphi)$, where $\varphi$ is a normal faithful state on $\cM$ (for details  see \cite{AlRZ}).
However, as we wish to put some emphasis on the universality of quantization,
we prefer to follow the Banach space theory approach developed by Dodds, Dodds and de Pagter \cite{DDdP} .

Given an element $f \in \widetilde{\cM}$ and $t \in [0, \infty)$, the generalised singular 
value $\mu_t(f)$ is defined by $\mu_t(f) = \inf\{s \geq 0 : \tau(\I - e_s(|f|)) \leq t\}$ 
where $e_s(|f|)$ $s \in \mathbb{R}$ is the spectral resolution of $|f|$. The function $t \to 
\mu_t(f)$ will generally be denoted by $\mu(f)$. For details on the generalised singular values 
see \cite{FK}.  (This directly extends classical notions where for any $f \in L^0_{\infty}{}$, 
the function $(0, \infty) \to [0, \infty] : t \to \mu_t(f)$ is known as the decreasing 
rearrangement of $f$.) We proceed to briefly review the concept of a Banach Function Space of 
measurable functions on $(0, \infty)$. (Necessary background is given in \cite{DDdP}.) A function norm 
$\rho$ on $L^0(0, \infty)$ is defined to be a mapping $\rho : L^0_+ \to [0, \infty]$ satisfying
\begin{itemize}
\item $\rho(f) = 0$ iff $f = 0$ a.e.  
\item $\rho(\lambda f) = \lambda\rho(f)$ for all $f \in L^0_+, \lambda > 0$.
\item $\rho(f + g) \leq \rho(f) + \rho(g)$ for all .
\item $f \leq g$ implies $\rho(f) \leq \rho(g)$ for all $f, g \in L^0_+$.
\end{itemize}
Such a $\rho$ may be extended to all of $L^0$ by setting $\rho(f) = \rho(|f|)$, in which case 
we may then define $L^{\rho}(0, \infty) = \{f \in L^0(0, \infty) : \rho(f) < \infty\}$. If 
now $L^{\rho}(0, \infty)$ turns out to be a Banach space when equipped with the norm 
$\rho(\cdot)$, we refer to it as a Banach Function space. If $\rho(f) \leq \lim\inf_n(f_n)$ 
whenever $(f_n) \subset L^0$ converges almost everywhere to $f \in L^0$, we say that $\rho$ 
has the Fatou Property. If less generally this implication only holds for $(f_n) \cup \{f\} 
\subset L^{\rho}$, we say that $\rho$ is lower semi-continuous. If further the situation $f 
\in L^\rho$, $g \in L^0$ and $\mu_t(f) = \mu_t(g)$ for all $t > 0$, forces $g \in L^\rho$ and 
$\rho(g) = \rho(f)$, we call $L^{\rho}$ rearrangement invariant (or symmetric). Using the 
above context Dodds, Dodds and de Pagter \cite{DDdP} formally defined the noncommutative space 
$L^\rho(\widetilde{\cM})$ to be $$L^\rho(\widetilde{\cM}) = \{f \in \widetilde{\cM} : \mu(f) \in 
L^{\rho}(0, \infty)\}$$and showed that if $\rho$ is lower semicontinuous and $L^{\rho}(0, 
\infty)$ rearrangement-invariant, $L^\rho(\widetilde{\cM})$ is a Banach space when equipped 
with the norm $\|f\|_\rho = \rho(\mu(f))$. 

Now for any Orlicz function $\phi$, the Orlicz 
space $L^\phi(0, \infty)$ is known to be a rearrangement invariant Banach Function space 
with the norm having the Fatou Property, see Theorem 8.9 in \cite{BS}. Thus on selecting $\rho$ to be 
 $\|\cdot\|_\phi$, the very general framework of Dodds, 
Dodds and de Pagter presents us with an alternative approach to realising noncommutative 
Orlicz spaces.

Note that this approach canonically contains the spaces of Kunze \cite{Kun} . To see this we recall that any Orlicz function is in fact continuous, non-negative and increasing 
on $[0, b_\phi)$. The fact that Kunze's approach to noncommutative Orlicz spaces is 
canonically contained in that of Dodds et al, therefore follows from the observation that if 
$b_\phi = \infty$, then for any $\lambda > 0$ and any $f \in \widetilde{\cM}$, we have  
$$\tau(\phi(\frac{1}{\lambda}|f|)) = \int_0^\infty \phi(\frac{1}{\lambda}\mu_t(|f|))\, 
\mathrm{d}t$$by \cite[2.8]{FK}. More generally we have the following lemma \cite{LM} :

\begin{lem} Let $\phi$ be an Orlicz function and $f \in \widetilde{\cM}$ a $\tau$-measurable element. Extend $\phi$ to a function on $[0, \infty]$ by setting $\phi(\infty) = \infty$. If $\phi(f) \in \widetilde{\cM}$, then $\phi(\mu_t(f)) = \mu_t(\phi(|f|)$ for any $t \geq 0$, and $\tau(\phi(|f|)) = \int_0^\infty \phi(\mu_t(|f|))\, 
\mathrm{d}t$.
\end{lem}

It is worth pointing out that the above lemma allows for the possibility that $a_\phi > 0$ and/or $b_\phi < \infty$. It is not difficult to see that if $a_\phi > 0$, then $$\cM \subset \{f \in \widetilde{\cM} :\phi(\lambda|f|) \in L_1(\cM, \tau) \quad \mbox{for some} \quad \lambda = 
\lambda(f) > 0\}.$$Thus this lemma is not contained in results like Remark 3.3 of 
\cite{FK}, which only hold for those elements of $\widetilde{\cM}$ for which $\lim_{t\to\infty}\mu_t(f) = 0$.

 Consequently, let us 
 take any Orlicz function $\phi$. Then the Orlicz space $L^{\phi}(0,\infty)$ is a Banach function space with a ``good'' norm.
Thus
\begin{equation}
\label{norma}
||f||_{\phi} = \inf \{ \lambda >0: \int_0^{\infty}dt \phi(\frac{\mu_t(f)}{\lambda}) \leq 1 \}
\end{equation}
gives the ``quantum'' Orlicz norm, where $f \in \widetilde{\cM}$.

In the next Theorem we collect our results on monotonicity of quantum maps with respect to the Orlicz norm given by the formula (\ref{norma}) (proofs will appear in \cite{LM}). However, we need some preliminaries. Firstly, following Arveson \cite{Ar} , we say that a completely positive map $T: \cM \to \cM$ is pure if, for every completely positive map $T^{\prime}: \cM \to \cM$, the property ``$T - T^{\prime}$ is a completely positive map'' implies that $T^{\prime}$ is a scalar multiple of $T$.   
Finally, a Jordan $*$-morphism $J : \cM \to \cM$ is $\epsilon-\delta$ absolutely 
continuous on the projection lattice of $\cM$ with respect to the trace $\tau$ \cite{L1}, if for any 
$\epsilon > 0$ there exists a $\delta > 0$ such that for any projection $e \in \cM$ we have 
$\tau(J(e)) < \epsilon$ whenever $\tau(e) < \delta$). We have

\begin{thm}
\label{twier}
Let $T : \cM \to \cM$ be a linear positive unital map. Then
\begin{equation}
\label{lala}
||T(f)||_{\phi} \leq C ||f||_{\phi}
\end{equation}
where $C$ is a positive constant, if 
\begin{enumerate}
\item{} $T$ is an inner automorphism, e.g. Hamiltonian type dynamics satisfying Borchers conditions (for exposition on Borchers conditions see e.g. Bratteli, Robinson book \cite{BR}) .
\item{} $T(\cdot) = \sum_1^{N < \infty}W_i^* \cdot W_i$ with $W_i \in \cM$.
\item{} $T(\cdot)$ is a pure unital normal CP map.
\item{} $T$ is a $\epsilon$-$\delta$ continuous normal Jordan morphism such that $\tau \circ J \leq \tau$. 
\end{enumerate} 
\end{thm}

The main idea of the proof is to show that generalized singular values $\mu_t(\cdot)$ are monotonic with respect to the maps $T$.
The rest of the proof follows from the definition of the Orlicz norm (\ref{norma}) and the monotonicity of the Orlicz function.

This theorem ensures the existence of extensions to quantum Orlicz space of a map $T:\cM \to \cM$ satisfying 
any of the conditions listed in Theorem \ref{twier}.
Consequently, we get the
 promised possibility of describing quantum dynamical system in terms of Quantum Orlicz spaces; so also in $L_p$-spaces!
 This explains why one can expect that dynamical maps defined
 for quantum $L_p$ spaces may have nice generalizations.

\section{Acknowledgments}

The support of Poland-South Africa Cooperation Joint project and (WAM) the support of the grant BW/5400-5-0307-7 is gratefully acknowledged.


\end{document}